\documentclass[a4paper]{jpconf}
\usepackage{graphicx}
\usepackage{amsmath}
\usepackage{amssymb}
\usepackage{array}
\usepackage{pifont}
\usepackage{tabularx}
\usepackage{latexsym}

\usepackage[usenames,dvipsnames]{xcolor}

\begin{document}
\title{Lepton Mixing Predictions from (Generalised) CP and Discrete Flavour Symmetry}

\author{Thomas Neder}

\address{School of Physics and Astronomy, University of Southampton, Southampton, SO17 1BJ, U.K.}

\ead{t.neder@soton.ac.uk}

\begin{abstract}
An important class of flavour groups, that are subgroups of $U(3)$ and that predict experimentally viable lepton mixing parameters including Majorana phases, is the $\Delta(6n^2)$ series. The most well-known member is $\Delta(24)=S_4$. I present results of several extensive studies of lepton mixing predictions obtained in models with a $\Delta(6n^2)$ flavour group that preserve either the full residual $Z_2\times Z_2$ or a $Z_2$ subgroup for neutrinos and can include a generalised CP symmetry. Predictions include mixing angles and Dirac CP phase generally; and if invariance under a generalised CP symmetry is included, also Majorana phases. For this, the interplay of flavour group and generalised CP symmetry has to be studied carefully. 
\end{abstract}

\section{Introduction}
This presentation concerns itself with two concepts that at a first glance seem to be entirely separate, whose interplay however is actually non-trivial and rather interesting. 

The first concept is the idea of promoting simultaneous conjugation of charge and parity of fields (CP) to a symmetry of a model: This means that at some high energy one requires the Lagrangian of the model not to change when every field is replaced by its CP-conjugate. In the Standard Model (SM), this is not the case and the question is whether this violation of CP-invariance happens spontaneously. To study this question, one needs to look at models that are explicitly invariant under CP conjugation at some high energy. In the work presented here, CP is added to the symmetries of the Standard Model in the unbroken state. However, it will not be discussed how exactly the breaking of CP invariance happens, because it is possible to determine the possible low-energy predictions  independently of the breaking mechanism. Generally, these predictions would be quite general, but in the interplay with discrete flavour symmetries they become discrete (i.e.\ predictive) and specific. In particular will it be possible to obtain predictions for the values of all mixing parameters, including the Majorana phases.

In the Standard Model, the three generations (or flavours, or families) of fermions are completely independent and are merely copies of each other. Flavour symmetries \cite{Altarelli:2010gt,King:2013eh} provide a link between the three generations by assigning them to representations of a flavour group in one way or another and demanding that the Lagrangian remains unchanged when fields are transformed according to their representation under the flavour group. In this article, only left-handed leptons are considered because this is already sufficient to arrive at low energy predictions for their mixing matrix. Furthermore, the three doublets of leptons are assigned to a 3-dimensional irreducible representation of the flavour group $G_F$. Thus in the unbroken state the theory is to be invariant under all of the following transformations allowed by $G_F$
\begin{equation}
\begin{pmatrix}(e,\nu_e)\\(\mu,\nu_\mu)\\(\tau,\nu_\tau)\end{pmatrix}\mapsto \rho(g) \begin{pmatrix}(e,\nu_e)\\(\mu,\nu_\mu)\\(\tau,\nu_\tau)\end{pmatrix}
\label{flavour_unbroken}
\end{equation}
where $\rho(g)$ is a $3\times 3$ matrix that correponds to group elements $g$ of the flavour group $G_F$ in the representation under which the three generations of leptons transform. The invariance under the flavour group will eventually need to be broken and from breaking it differently for charged leptons and neutrinos the mixing matrix will arise. Again independently of the actual mechanism, imagine therefore that $G_F$ is broken to different subgroups, namely $G_\nu$ for neutrinos and $G_e$ for charged leptons: The Lagrangian will be invariant under the following transformations in the broken state
\begin{equation}
\begin{pmatrix}e\\ \mu\\ \tau\end{pmatrix}\mapsto \rho(g_e) \begin{pmatrix}e\\ \mu\\ \tau\end{pmatrix} \text{ and } \begin{pmatrix}\nu_e\\ \nu_\mu\\ \nu_\tau\end{pmatrix}\mapsto \rho(g_\nu) \begin{pmatrix}\nu_e\\ \nu_\mu\\ \nu_\tau\end{pmatrix}
\label{Eq2}
\end{equation}
where now $\rho(g_e)$ and $\rho(g_\nu)$ are $3\times3$-matrices that correspond to elements of the subgroups that remain unbroken but only in their sector, $G_e$ and $G_\nu$ respectively, again in the same representation as in the unbroken state.

After having defined flavour symmetries above, the same needs to and will be done for CP transformations in the next section. There, it will also be discussed why one has to be careful how to define CP transformations in the presence of several generations of fermions and in particular if the theory is to be invariant under a flavour symmetry. This leads to a consistency requirement which CP and flavour transformations have to obey if not one is to accidentally violate invariance under the other.

In the section thereafter, it will be explained how predictions for the mixing matrix can be obtained from this symmetry-based approach in the different types of models with flavour symmetries. For the most constrained and predictive class of models with flavour symmetries, only $\Delta(6n^2)$ groups remain viable flavour symmetry groups.\footnote{The $\Delta(6n^2)$ group are not always the minimal flavour group for a certain mixing pattern, however, all candidate groups are subgroups of a $\Delta(6n^2)$ group and the 3-dimensional irreducible representations are inherited by $\Delta(6n^2)$.} After that, the predictions for mixing parameters including Majorana phases will be presented for several classes of models employing $\Delta(6n^2)$ groups with one free unitary rotation, respectively. It will turn out that despite the free continuous parameters, the predictions are rather constrained.

In the final section, conclusions are drawn and possible directions for future work are indicated.

\section{(Generalised) CP Transformations}
The kind of CP transformations that are sometimes called canonical CP transformations act on a scalar field  $\phi(x)$ in the following way:
\begin{equation}
\phi(x)\mapsto \phi^{CP}=e^{i \varphi}\phi^\ast(x^P)
\end{equation}
where $\varphi$ is a phase that might depend on the kind of field and $x^P=(t,-{\bf x})$ is the parity-conjugated spacetime coordinate. But now imagine that several copies of the field $\phi$ exist in such a way that the Lagrangian allows for unitary rotations transforming these copies into each other. If all of the copies of $\phi$ would still transform under CP in the canonical way, then one could think that the CP properties of the set of copies would be basis-dependent, which is unphysical. However, this argument is wrong, because the unitary rotations mix the different states in an arbitrary way which is incompatible with canonical CP transformations acting diagonally on the copies of fields.
So already the existence of several copies of a field that allow changes of basis demands that a CP transformation allows for these changes of basis \cite{Grimus:1995zi,Feruglio:2012cw}. This can be achieved by promoting the phase $e^{i\varphi}$ to a unitary matrix, here called $X$, where now $\Phi$ is the multiplet of the copies of the field $\phi$:
\begin{equation}
\Phi(x)\mapsto \Phi^{CP}=X \Phi^\ast(x^P).
\label{gcp}
\end{equation}
Indeed it can be shown, that in situations similar to the one above, canonical CP can be violated while observables of CP show no CP violation because generalised CP is conserved correctly \cite{Chen:2014tpa}. This means that in such a situation the correct definition of CP is the generalised one because it accommodates for changes of basis. The above applies exactly in the same way to several copies of fermions because in CP transformations of fermions only an additional matrix appears that acts on spinor indices and not among copies of the fields and thus can be neglected without loss of generality.

This situation is complicated further if the theory already is invariant under some symmetry group that acts on the same multiplets that allow for generalised CP transformations. In this case flavour and CP symmetry transformations need to fulfill a consistency condition \cite{Holthausen:2012dk,Chen:2014tpa,Feruglio:2012cw}: Assume that the Lagrangian is invariant under flavour transformations like in Eq.(\ref{flavour_unbroken}) of a flavour group $G_F$. A matrix $X$ can appear in a generalised CP transformation acting on the multiplet of leptons if $G_F$ contains elements $g$ and $g'$ such that the following equation is fulfilled
\begin{equation}
 X^\dagger \rho^\ast(g) X=\rho(g')
 \label{consistency}
\end{equation}
where $\rho(g)$ and $\rho(g')$ are the $3\times 3$-matrices that correspond to $g$ and $g'$ in the representation of the leptons. Because the symmetry of the model is enhanced by the flavour symmetry, matrices $X$ can appear that are connected to the identity not only by basis transformations but also by appropriate flavour transformations.


For the groups considered in this paper, the $X$ matrices allowed in CP transformations are proportional to the representation matrices of the flavour group \cite{Chen:2014tpa,King:2014rwa}:
\begin{equation}
 \{X\}=e^{i \alpha}\rho(G)
\end{equation}
with an additional phase $\alpha$ that will cancel in physical predictions but is shown here for completeness. 

When the flavour group $G_F$ is broken to its subgroups $G_e$ and $G_\nu$ in the different sectors, then invariance under CP can be spontaneously broken as well. This follows from the fact that $X$ matrices that are consistent with $G_e$ or $G_\nu$ in one sector respectively are not necessarily consistent in the other sector. 
\section{Residual Symmetries and Flavour Symmetries}
Consider a multiplet of Majorana or Dirac fields $\Psi$ with projections on left/right-handed components $\Psi_{L/R}$ that have the following mass terms after all symmetry breakings have occured: 
\begin{equation}
\mathcal{L}_\text{Majorana}=\Psi_L^T M_\nu \Psi_L \text{ or } \mathcal{L}_\text{Dirac}=\Psi_L^\dagger M_e \Psi_R+\text{h.c.}
\end{equation}
Invariance under the transformations in Eq.(\ref{Eq2}) puts the following constraints on the mass matrices
\begin{equation}
\rho(g_\nu)^TM_\nu\rho(g_\nu)=M_\nu\text{ and } \rho(g_e)^\dagger M_e M_e^\dagger \rho(g_e)=M_e M_e^\dagger,
\end{equation}
whereas from invariance under transformations as in Eq.(\ref{gcp}) follows
\begin{equation}
X^TM_\nu X=M^\ast_\nu\text{ and } X^\dagger M_e M_e^\dagger X=M_e^\ast M_e^T.
\end{equation}

Both Majorana and Dirac mass matrices can be diagonalised by  appropriate unitary matrices $U_{e,L/R}=(u^e_1,u^e_2,u^e_3)$ or $U_\nu=(u^\nu_1,u^\nu_2,u^\nu_3)$ respectively, where the lower case $u_i^{\nu/e}$ denote the columns of the matrices.
The convention chosen is that the mixing matrices act actively on the multiplets of fields such that the physical mixing matrix that appears in the vertex of W-boson, neutrino, and antifermion becomes $U_\text{PMNS}=U_{e,L}^\dagger U_\nu$.
For Majorana fermions one can define the following matrices:
\begin{equation}
 G_1=+u^\nu_1 (u^\nu_1)^\dagger-u^\nu_2 (u^\nu_2)^\dagger-u^\nu_3 (u^\nu_3)^\dagger,
 \label{Gres1}
\end{equation}
\begin{equation}
 G_2=-u^\nu_1 (u^\nu_1)^\dagger+u^\nu_2 (u^\nu_2)^\dagger-u^\nu_3 (u^\nu_3)^\dagger,
 \label{Gres2}
\end{equation}\begin{equation}
 G_3=-u^\nu_1 (u^\nu_1)^\dagger-u^\nu_2 (u^\nu_2)^\dagger+u^\nu_3 (u^\nu_3)^\dagger.
 \label{Gres3}
\end{equation}

Together with the identity matrix these form a $Z_2\times Z_2$ group. For 3 massive and non-degenerate neutrinos this is the maximal residual symmetry in the following sense: The subgroup $G_\nu$ of $G_F$ that is conserved by the Majorana mass term needs to be a subgroup of a $Z_2\times Z_2$ group (or equal to it) or the constraints imposed on $M_\nu$ from the invariance under $G_\nu$ force neutrinos to be degenerate or even massless.\footnote{Technically one could extend the residual symmetry by the negative identity matrix to $Z_2\times Z_2\times Z_2$ but this would not impose any further constraints on the mixing matrix.} 

Depending on how much of this residual symmetry is identified with a subgroup of $G_F$, models can be classified and will have different properties: Firstly, models where the whole residual $Z_2\times Z_2$ is identified with $G_\nu$ are called direct models. In this case, all 3 $G_i$ are known in a certain basis and the mixing matrix of the neutrinos is fixed in that basis, even including the Dirac phase but not including the Majorana phases. The reason for this is that the $i$th column of the mixing matrix is an eigenvector of $G_i$ with eigenvalue $+1$:
\begin{equation}
 G_i u^\nu_i=+u^\nu_i.
\end{equation}
The Majorana phases remain free because they are just overall phases on the columns.
In this case, all mixing matrices that are allowed in a $G_F$-symmetric model can be obtained by listing all $Z_2\times Z_2$ subgroups of $G_F$.
If on the other hand $G_\nu\simeq Z_2$ and $G_\nu$ is identified with one of the $Z_2$ factors of the residual $Z_2\times Z_2$, a model would be called semidirect and only one of the columns of the mixing matrix is determined from symmetry as an eigenvector of the corresponding $G_i$.
Again, it is possible to list all $Z_2$ subgroups of $G_F$. Every $Z_2$ subgroup gives rise to a class of mixing matrices that can be parametrized as
\begin{equation}
 U=U_{Z_2}U_{2\times2}
 \label{semidirect1}
\end{equation}
where $U_{Z_2}$ is a unitary matrix with one of the three columns fixed as an eigenvector of the $G_i$ coming from $G_F$ and the other columns arbitrary but fixed throughout the analysis. $U_{2\times2}$ is a unitary matrix that rotates the two free columns and e.g.\ in case of having the first column of $U_{Z_2}$ determined by $Z_2\subset G_F$ has the form
\begin{equation}
 U_{2\times 2}=\begin{pmatrix}1&0&0\\0&\cos(\theta)&\sin(\theta)e^{i\varphi}\\0&-\sin(\theta)e^{-i\varphi}&\cos(\theta)\end{pmatrix}
 \label{semidirect2}
\end{equation}
where $\theta$ and $\phi$ are real parameters and analogous for the other columns.

If the flavour group $G_F$ is broken completely in the sector of neutrinos, $G_\nu\simeq \{1\}$, then no constraints on the mixing matrix follow from the invariance under $G_F$ in this immediate way from symmetry but can be obtained in a different way \cite{King:2013eh}. In this work, direct and semidirect models are considered.

What is peculiar about the residual symmetry is that it is an accidental symmetry that always exists as defined by Eqs.(\ref{Gres1}-\ref{Gres3}) independently of it being incorporated into a bigger group.\footnote{Although also semidirect models are considered in this work, this could lead to the conjecture that if it is impossible to identify the residual symmetry with a subgroup of $G_F$ that one has chosen the wrong $G_F$ or that it is simply not big enough. In a way, direct models thus seem more natural, because all accidental symmetries are accounted for.} 

Similar to the way described in this section so far, also the mass term of a multiplet of Dirac fermions has a maximal residual symmetry which forms a $U(1)\times U(1)$ group\footnote{Again, one could add another $U(1)$ which corresponds to an overall phase on the mass matrix which will not impact the subsequent discussion.}
\begin{equation}
 \left\{U_e\begin{pmatrix}e^{i \alpha}&0&0\\0&e^{i \beta}&0\\0&0&e^{-i\alpha-i\beta}\end{pmatrix}U_e^\dagger:\alpha,\beta\in[0,2\pi)\right\}\simeq U(1)\times U(1).
\end{equation}
In particular, $U(1)\times U(1)$ contains all groups of the kind $Z_p\times Z_q$ as subgroups with $q,p$ natural numbers. If the form of one of the $Z_q$ factors is known from $G_F$ and $q\geq3$, then the mixing matrix can be completely determined as the eigenvectors of the generator of this $Z_q$ group. Only if $q=2$, then similarly to the semidirect case for Majorana fermions, only one column of the matrix $U_e$ is fixed. The latter case will be called charged-lepton-semidirect and in this case the mixing matrix of the charged leptons, $U_e$, will contain a free unitary rotation as in Eqs.(\ref{semidirect1}) and (\ref{semidirect2}). This is interesting because in contrast to the neutrino-semidirect case, in the physical mixing matrix the free rotation will act on rows instead of columns which opens the field for new mixing predictions as will be seen in the discussion of semidirect models.

The discreteness of the residual symmetry for Majorana neutrinos could be seen as a motivation for considering discrete flavour groups $G_F$: Generally (and as will be shown for an example later) continuous groups allow for an infinite number of $Z_2\times Z_2$ subgroups and thus a discrete and specific prediction for the mixing matrix from symmetry would become impossible. On the other hand, Dirac fermions would be perfectly happy with a continuous residual symmetry.

Many discrete groups were considered as candidates for $G_F$ in direct, semidirect and indirect models, often motivated by the conjecture that the reactor neutrino mixing angle is zero, $\theta_{13}=0$. After the measurement of a non-zero $\theta_{13}$ \cite{An:2012eh,Ahn:2012nd,Abe:2011fz}, many of these models were ruled out and searches for other groups started, in particular using the group theory program GAP \cite{GAP4:2011}. In two such searches \cite{d150lam,Holthausen:2012wt}, it was discovered that only groups that belong to the class of groups called $\Delta(6n^2)$ \cite{Escobar:2008vc} could provide columns that would remain viable in direct or semidirect models. Motivated by this finding all $\Delta(6n^2)$ groups were studied for their possible predictions in direct and semidirect models \cite{King:2014rwa,KingVNA,NederMXA,DingORA,Hagedorn:2014wha}. Later it was proved \cite{FonsecaKOA} by listing all possible mixing matrices in direct models for general finite groups, that only $\Delta(6n^2)$ groups and certain subgroups thereof can provide mixing matrices that remain experimentally viable. As mentioned earlier, the procedure of deriving the structure of mixing matrices from a partially broken flavour group cannot predict Majorana phases. On the other hand, by combining a flavour symmetry with a CP symmetry consistent with it, all phases of the mixing matrices can be calculated and are predicted to particular discrete values in a direct model or can additionally only depend on up to two parameters that are strongly constrained by measurements in semidirect models. Taking this as an incentive, the possible predictions for mixing matrices from $\Delta(6n^2)$ groups combined with their consistent CP transformations were examined in \cite{King:2014rwa,DingORA,Hagedorn:2014wha}. It is the results and consequences of these studies, starting from \cite{Escobar:2008vc}, that will be reported in the following.

\begin{figure}
\centering
\includegraphics[width=0.68\textwidth]{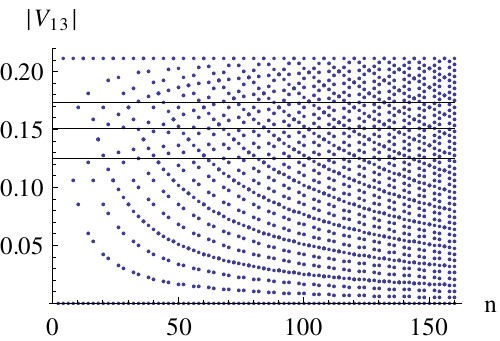}
\caption{The possible predictions for the value of $|U^\text{PMNS}_{13}|$ in direct models with $\Delta(6n^2)$ flavour groups. Each dot corresponds to a different choice $Z_2\times Z_2$ subgroup and the horizontal lines indicate the central value and three sigma range from \cite{Capozzi:2013csa}. When $3$ divides $n$, points are missing because for such $\Delta(6n^2)$ groups those subgroups are missing.}
\label{u13vsn}
\end{figure}

The groups $\Delta(6n^2)$ are isomorphic to a semidirect product
\begin{equation}
\Delta(6n^2)\simeq (Z_n\times Z_n)\rtimes S_3
\end{equation}
and are not as obscure as the name or notation suggests. An easy way to present these groups is in terms of four generators called $a,b,c,d$, where $a$ and $b$ generate the $S_3$ subgroup and $c$ and $d$ each generate one $Z_n$ subgroup. $c$ and $d$ commutate, while $a$ and $b$ have to fulfill the rules of cycling three objects and interchanging two:
\begin{equation}
a^3=b^2=(ab)^2=1
\end{equation}
The semidirect product, ``$\rtimes$'', imposes conditions on products of $a,b$ with $c,d$ \cite{Escobar:2008vc}. This presentation has the advantage that every group element $g\in \Delta(6n^2)$ can be expressed by the generators in the following way
\begin{equation}
 g=a^\alpha b^\beta c^\gamma d^\delta
\end{equation}
with $\alpha=0,1,2$; $\beta=0,1$; $\gamma,\delta=0,\ldots,n-1$.
In a way the groups can be thought of as two cyclic groups with $n$ elements with permutations of three objects acting on them. The $\Delta(6n^2)$ groups for small $n$ are well-known: $\Delta(6)=S_3$, $\Delta(24)=S_4$, $\Delta(54)$ and $\Delta(96)$.
Constraining the representation of lepton doublets to a 3-dimensional one, the $\Delta(6n^2)$ groups still have to offer $2n-1$ different representations. However, it can be shown that it is sufficient to only consider one of these representations \cite{KingVNA}. In the representation chosen without loss of generality for this analysis, the generators have the following matrix form:
\begin{equation}
a=\begin{pmatrix}0&1&0\\0&0&1\\1&0&0\end{pmatrix},b=-\begin{pmatrix}0&0&1\\0&1&0\\1&0&0\end{pmatrix},c=\begin{pmatrix}\eta&0&0\\0&\eta^{-1}&0\\0&0&1\end{pmatrix},d=\begin{pmatrix}1&0&0\\0&\eta&0\\0&0&\eta^{-1}\end{pmatrix}
\end{equation}
with $\eta=e^{2\pi i/n}$. To arrive at the mixing matrix predictions, first all $Z_2\times Z_2$ subgroups are listed as candidates for $G_\nu$. Concerning $G_e$, only $Z_3$ groups turn out to produce experimentally viable mixing matrices, although other candidates for $G_e$ had been considered \cite{KingVNA}. Knowing $G_\nu$ and $G_e$ in a direct model completely determines the mixing matrix including the Dirac CP phase except for the ordering of rows and columns. It is found that one column always is trimaximal, i.e.\ $\propto (1,1,1)^T$ and has to serve as middle column of the mixing matrix. Additionally, one requires that the entry with the smallest absolute value becomes $U_{13}^\text{PMNS}$. The only freedom that remains after that, is interchanging the second and third row, which corresponds to two different predictions for $\theta_{23}$. As the Majorana phases cannot be predicted using this symmetry approach, the effect of additionally imposing CP transformations consistent with $\Delta(6n^2)$ was examined. 
Out of these, only CP transformations that are consistent with $G_\nu$ can be imposed onto the neutrino mass matrix $M_\nu$, while analogously only CP transformations that are consistent with $G_e$ can be used to constrain the charged lepton mass matrix. After this procedure, in the PDG convention \cite{pdg} all experimentally viable mixing matrices have the form
\begin{equation}
 U_\text{PMNS}=\left(
\begin{array}{ccc}
 \sqrt{\frac{2}{3}} \cos \left(\frac{\pi  \gamma }{n}\right) & \frac{1}{\sqrt{3}} & 
   \sqrt{\frac{2}{3}} \sin \left(\frac{\pi  \gamma }{n}\right) \\
 -\sqrt{\frac{2}{3}} \sin \left(\pi  \left(\frac{\gamma }{n}+\frac{1}{6}\right)\right) & \frac{1}{\sqrt{3}} & \sqrt{\frac{2}{3}} \cos \left(\pi  \left(\frac{\gamma }{n}+\frac{1}{6}\right)\right)
   \\
 \sqrt{\frac{2}{3}} \sin \left(\pi  \left(\frac{1}{6}-\frac{\gamma }{n}\right)\right) & -\frac{1}{\sqrt{3}} &  \sqrt{\frac{2}{3}} \cos \left(\pi  \left(\frac{1}{6}-\frac{\gamma }{n}\right)\right)
   \\
\end{array}
\right)\begin{pmatrix}1&0&0\\0&[i]ie^{-i6\pi(\gamma+x)/n}&0\\0&0&[i]i\end{pmatrix}
\end{equation}
with $\gamma,x=0,\ldots,n-1$ and the parts in square brackets correspond to different possible choices of the Majorana phases. Each value of $\gamma$ correponds to a different $Z_2\times Z_2$ subgroup. In figure \ref{u13vsn} the possible predictions for the value of $|U^\text{PMNS}_{13}|$ in direct models with $\Delta(6n^2)$ flavour groups is shown. There, each dot corresponds to a different choice of $\gamma$ and thus a different $Z_2\times Z_2$ subgroup. When $3$ divides $n$, points are missing because for such $\Delta(6n^2)$ groups those subgroups are missing which in a way renders such groups more predictive.\footnote{The group with n=42 produces no predictions within the three sigma range, contrasting well-regarded hints in the literature \cite{Adams1979}.} In the limit $n\rightarrow \infty$, a prediction of the value of $\theta_{13}$ becomes impossible. As 
\begin{equation}
\Delta(6n^2)\simeq(Z_n\times Z_n)\rtimes S_3 \xrightarrow{n\rightarrow \infty}(U(1)\times U(1))\rtimes S_3,
\end{equation}
this constitutes an example for a continuous group that does not predict discrete and specific values of a parameter in a direct model.
For an arbitrary but fixed $\gamma$, it is currently not possible to determine the ordering of the second and third row from experiment. Thus for each value of $\theta_{13}$ two values of $\theta_{23}$ are predicted. This can also be expressed by the sum rule  
\begin{equation}
\theta_{23}=45^\circ \mp \theta_{13}/\sqrt{2}.
\end{equation}
A further important prediction is that the Dirac CP phase is zero: $\delta_{CP}=0$ whereas global fits are starting to show a slight preference for a non-zero value \cite{Tortola:2012te,Capozzi:2013csa,Gonzalez-Garcia:2014bfa}. To summarize, these predictions are general for all finite groups where the lepton-doublets transform in a 3-dimensional representation and where the flavour group is broken to a $Z_2\times Z_2$ subgroup for neutrinos. If the lepton mixing parameters are experimentally found to differ from the above predictions and corrections through renormalisation cannot account for the difference, then direct flavour models with finite groups are excluded as a paradigm.\footnote{In a way, as continuous groups will not produce specific predictions in a direct model, one could think that the most important motivation for considering discrete groups lies in the Majorana nature of neutrinos. If discrete groups fail to predict the correct mixing matrix one could thus be tempted to consider doubting the Majorana nature of neutrinos.}

The previous discussion exhausts the topic of direct models with discrete flavour groups. In the following we focus on the possible predictions for the lepton mixing matrix that can be obtained in semidirect models with a $\Delta(6n^2)$ flavour group. We consider breaking the flavour group in such a way that only one unitary rotation is free as in Eqs.(\ref{semidirect1}) and (\ref{semidirect2}). This will entail neutrino-semidirect breaking, where $G_\nu\simeq Z_2$ while the charged lepton mixing matrix is completely fixed, i.e.\ $G_e\simeq Z_2\times Z_2, Z_p$, as well as charged-lepton-semidirect breaking, where $G_e\simeq Z_2$ and $G_\nu\simeq Z_2 \times Z_2$. Each combination of $G_e$ and $G_\nu$ will provide one column of the physical mixing matrix in the neutrino-semidirect case and one row of the mixing matrix in the charged-lepton-semidirect case. In table 1 the fixed columns for each combination are shown for $G_\nu\simeq Z_2$. A tick mark denotes that this column is experimentally viable as one of the columns of the physical mixing matrix, possibly requiring reordering of rows of the matrix. In table 2 the predicted rows of the mixing matrix are shown for $G_e\simeq Z_2$, again not barring reordering.

\begin{figure}
 \includegraphics[width=0.32\textwidth]{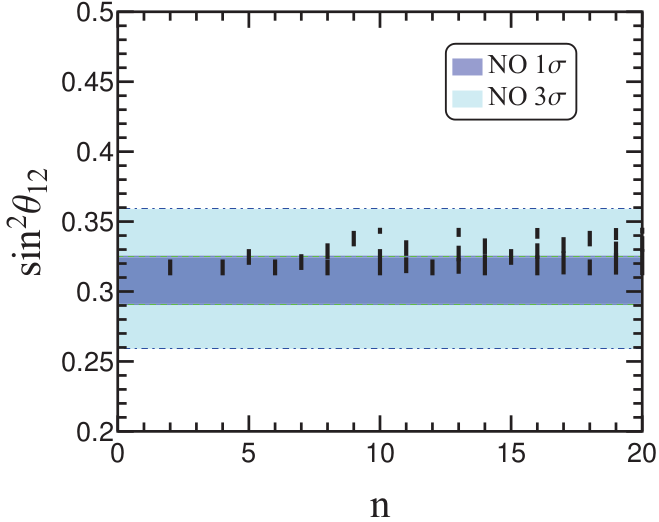}
 \includegraphics[width=0.32\textwidth]{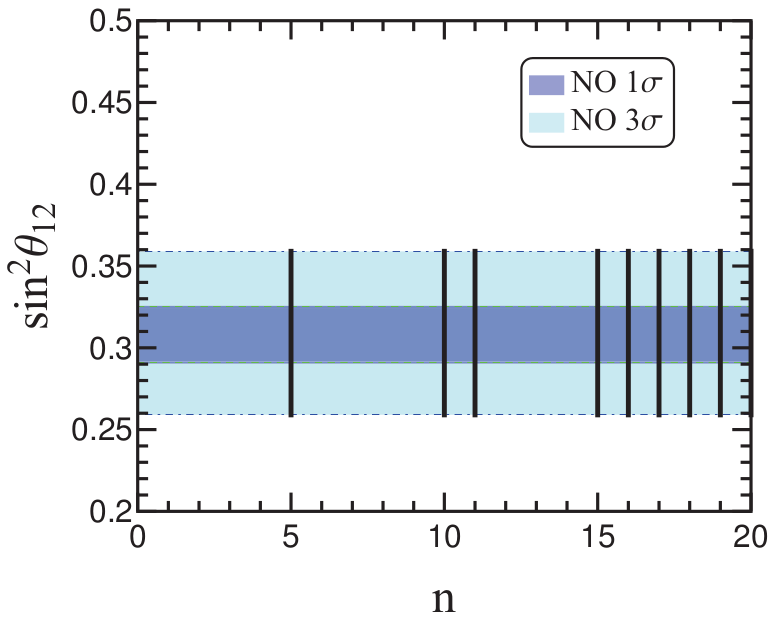}
 \includegraphics[width=0.32\textwidth]{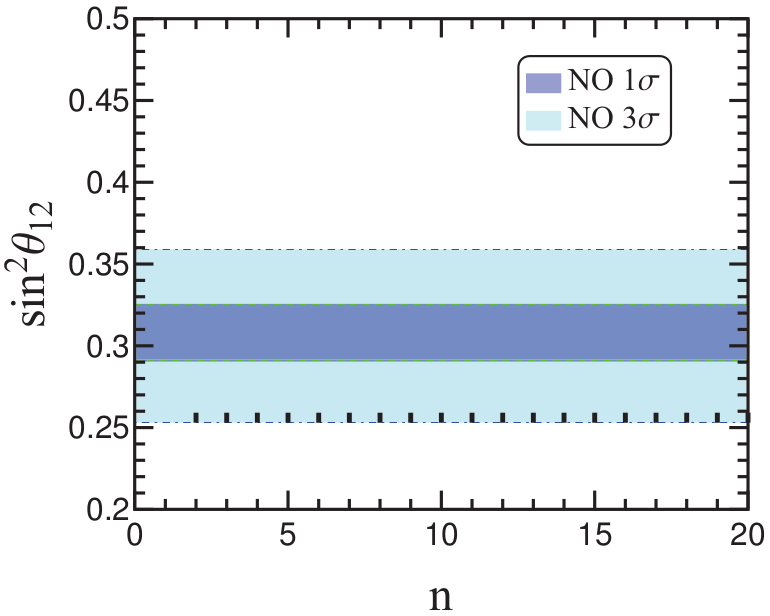}\\
 \includegraphics[width=0.32\textwidth]{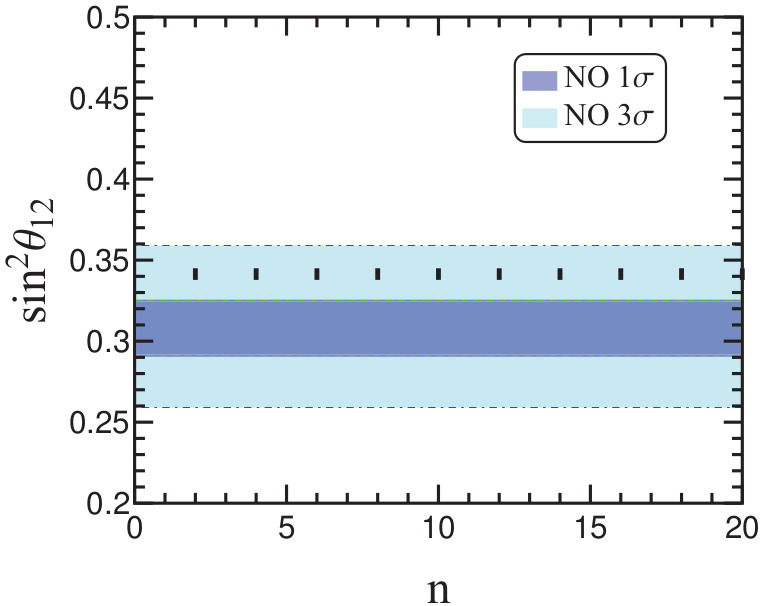}
 \includegraphics[width=0.32\textwidth]{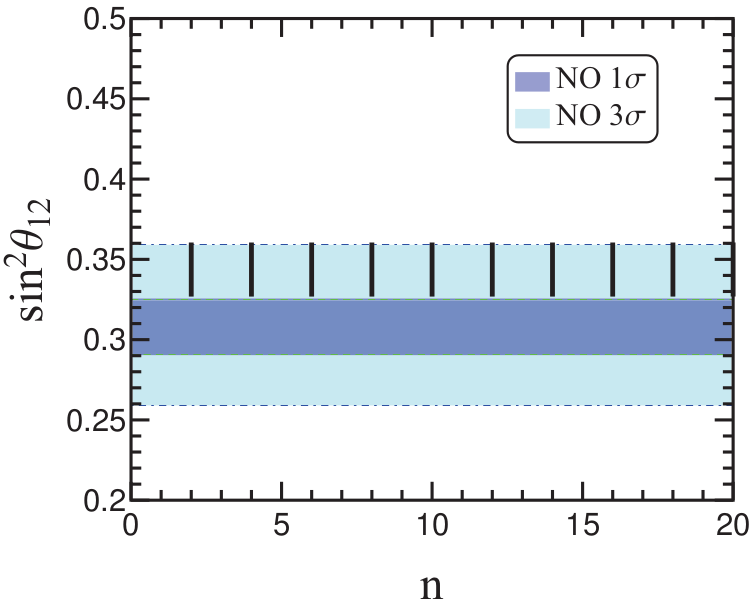}
 \label{semidirect_theta12}
 \caption{Predictions for $\sin^2\theta_{12}$ for the five classes of experimentally viable mixing matrices for semidirect models with a $\Delta(6n^2)$ flavour group. The coloured bands show the one and three sigma bands respectively.}
 \vspace{1cm}
 \includegraphics[width=0.32\textwidth]{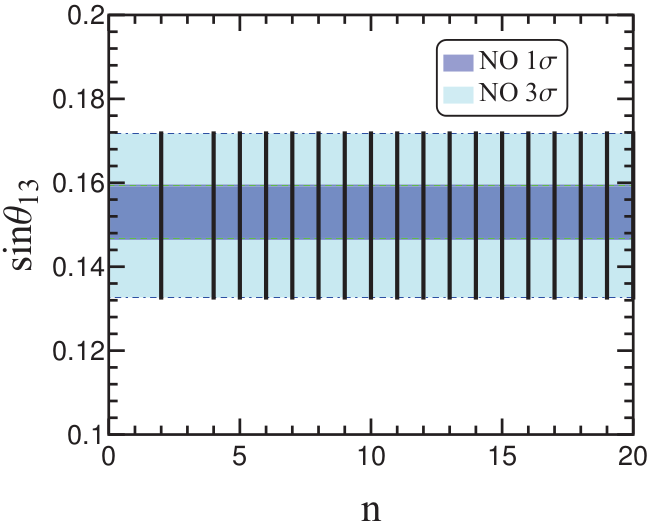}
 \includegraphics[width=0.32\textwidth]{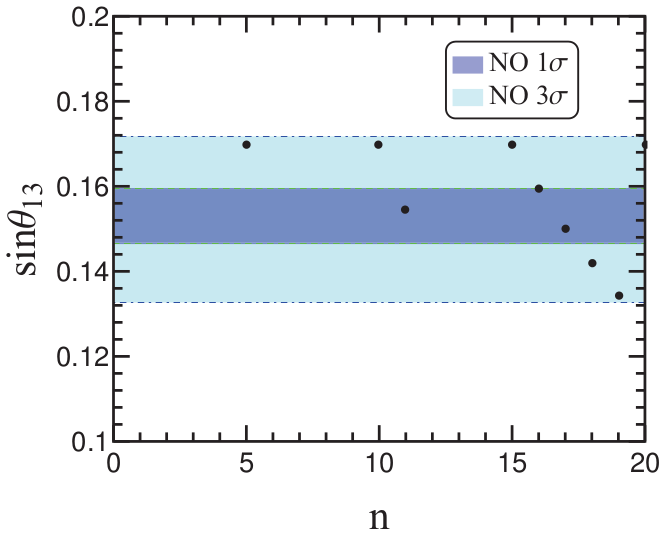}
 \includegraphics[width=0.32\textwidth]{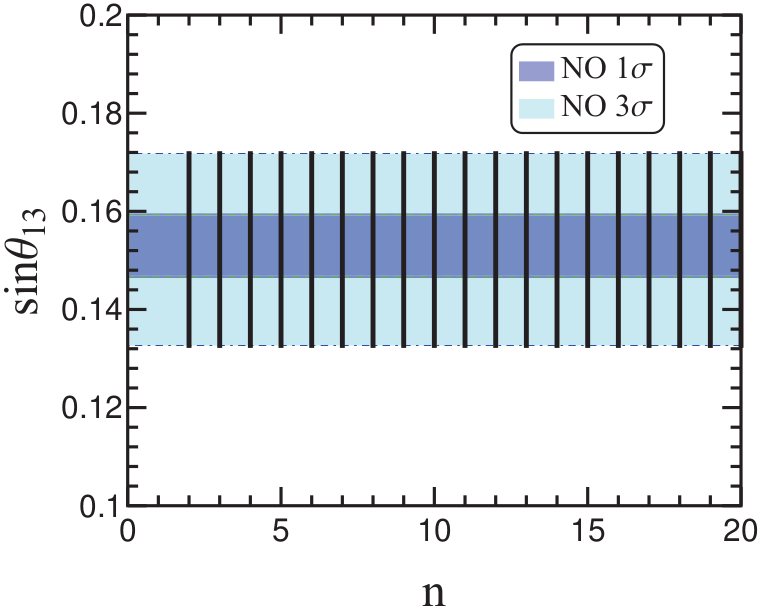}\\
 \includegraphics[width=0.32\textwidth]{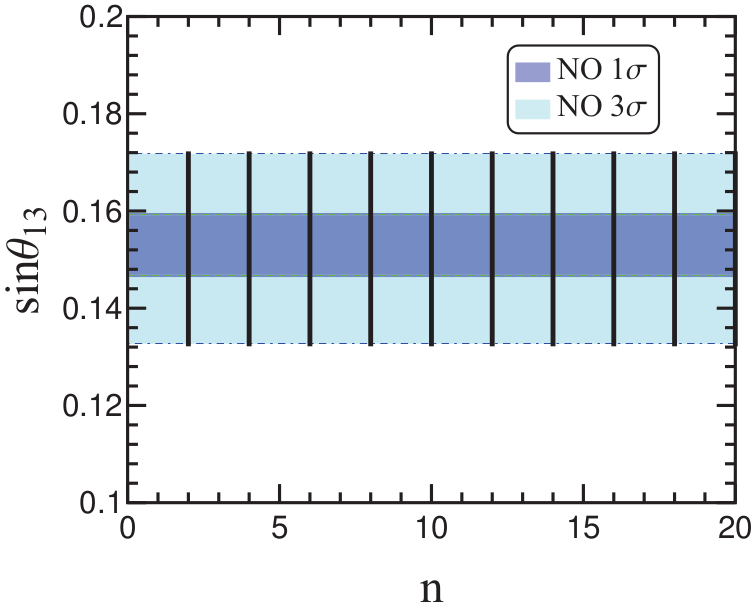}
 \includegraphics[width=0.32\textwidth]{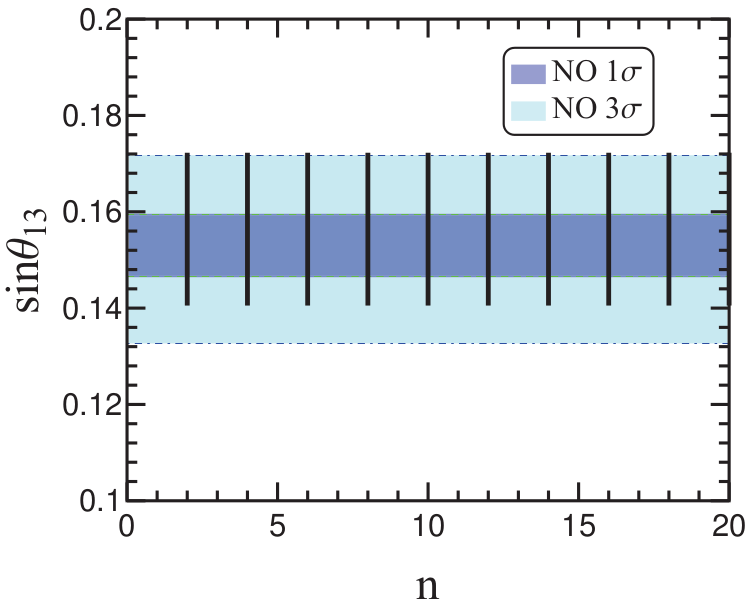}
 \caption{Predictions for $\sin\theta_{13}$ for the five classes of experimentally viable mixing matrices for semidirect models with a $\Delta(6n^2)$ flavour group. The coloured bands show the one and three sigma bands respectively.}
\end{figure}

\begin{figure}
 \includegraphics[width=0.32\textwidth]{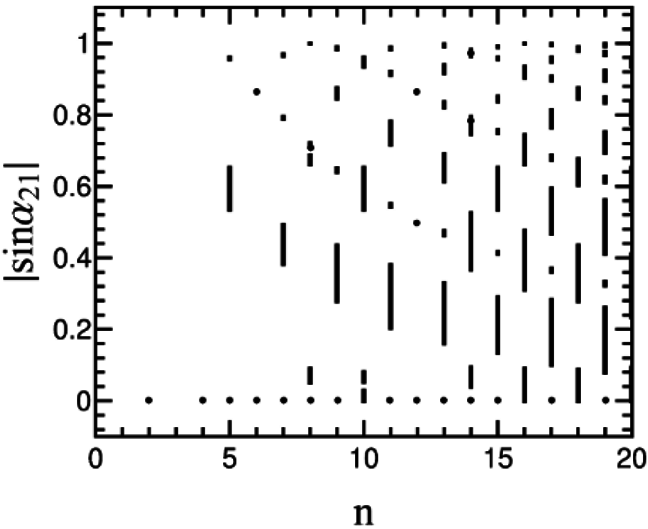}
 \includegraphics[width=0.32\textwidth]{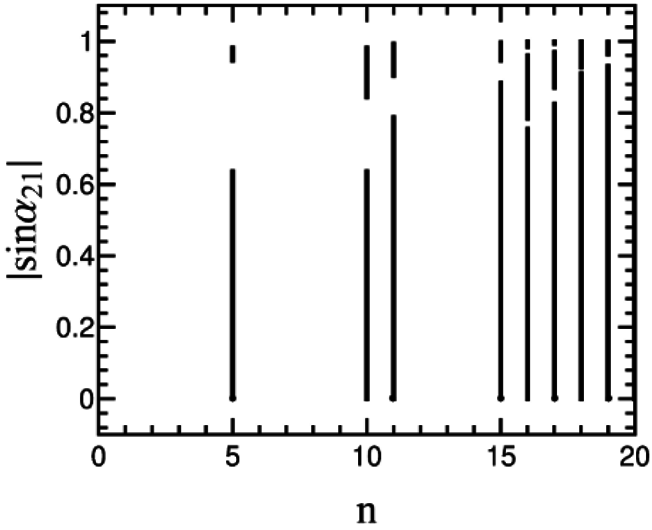}
 \includegraphics[width=0.32\textwidth]{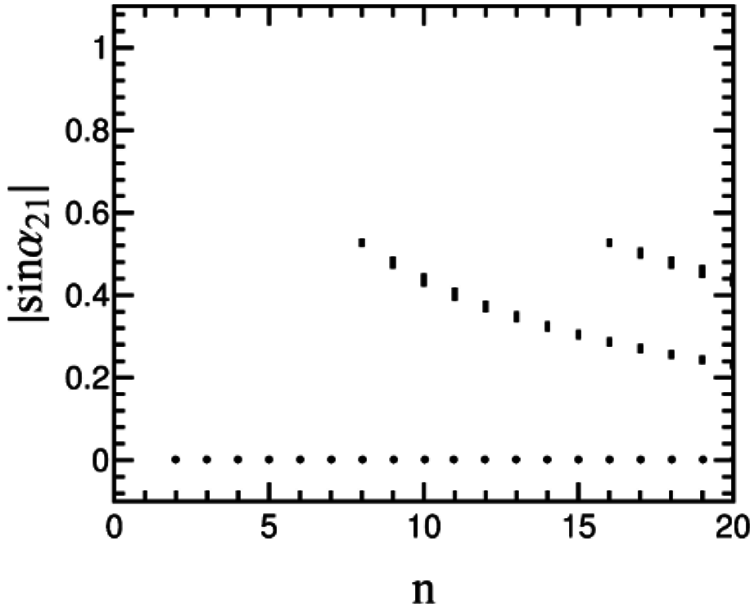}\\
 \includegraphics[width=0.32\textwidth]{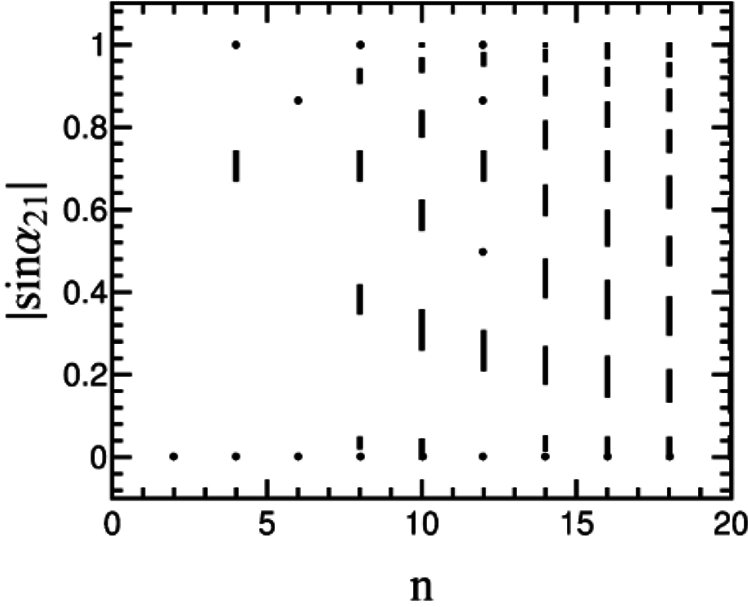}
 \includegraphics[width=0.32\textwidth]{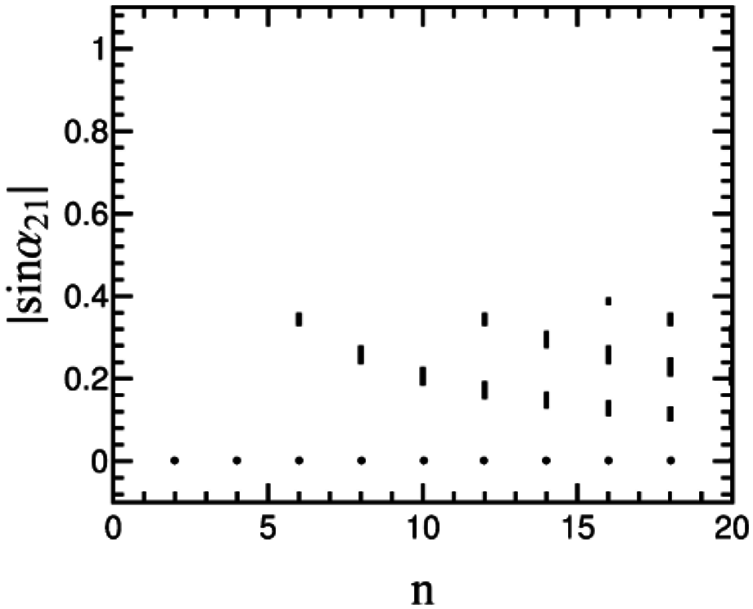}
 \caption{Predictions for $|\sin \alpha_{21}|$ for the five classes of experimentally viable mixing matrices for semidirect models with a $\Delta(6n^2)$ flavour group.}
 \vspace{1cm}
 \includegraphics[width=0.32\textwidth]{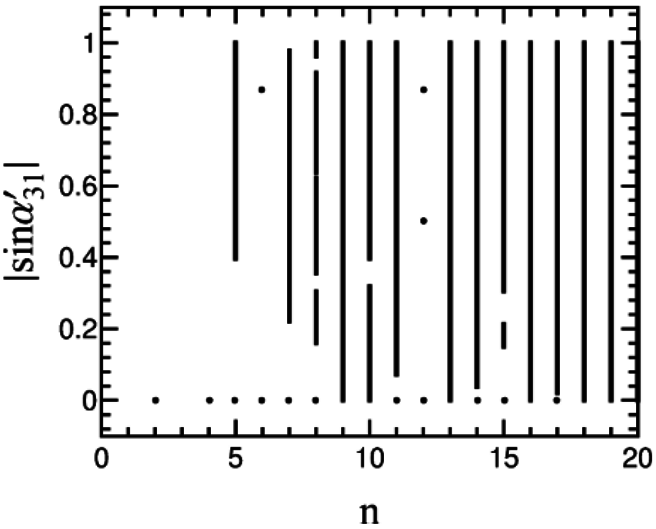}
 \includegraphics[width=0.32\textwidth]{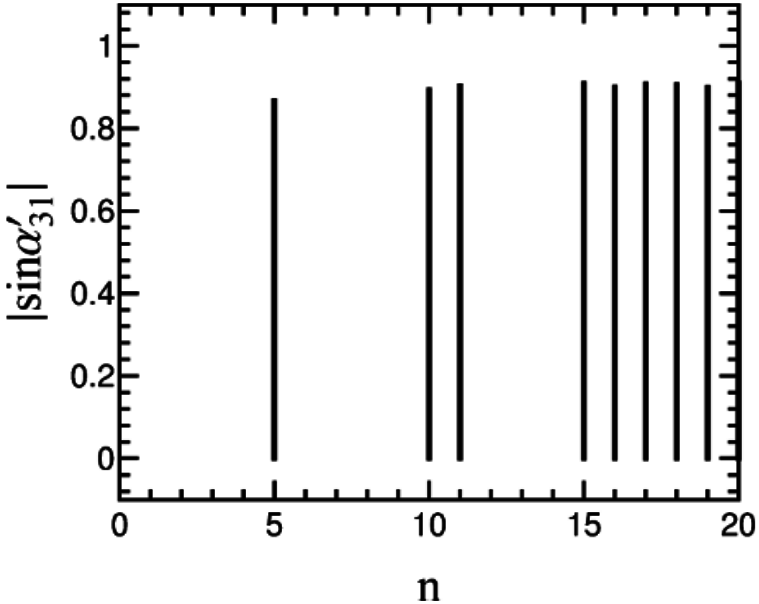}
 \includegraphics[width=0.32\textwidth]{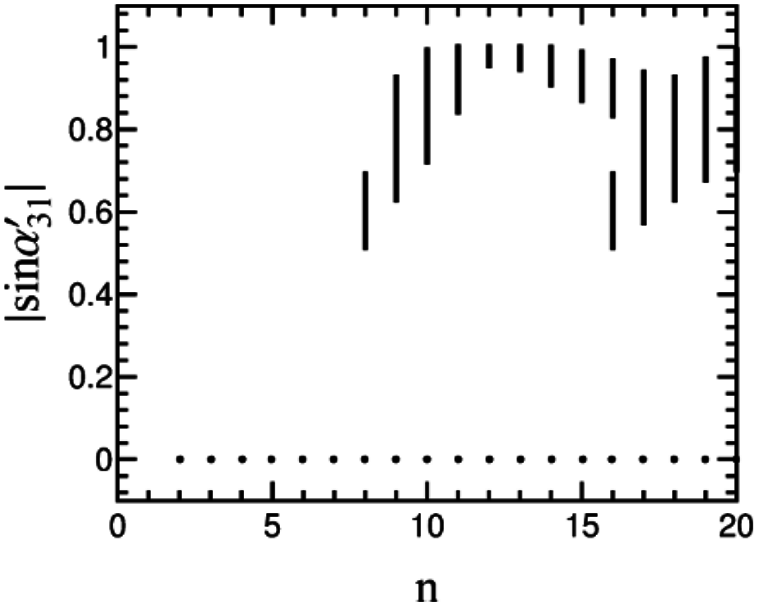}\\
 \includegraphics[width=0.32\textwidth]{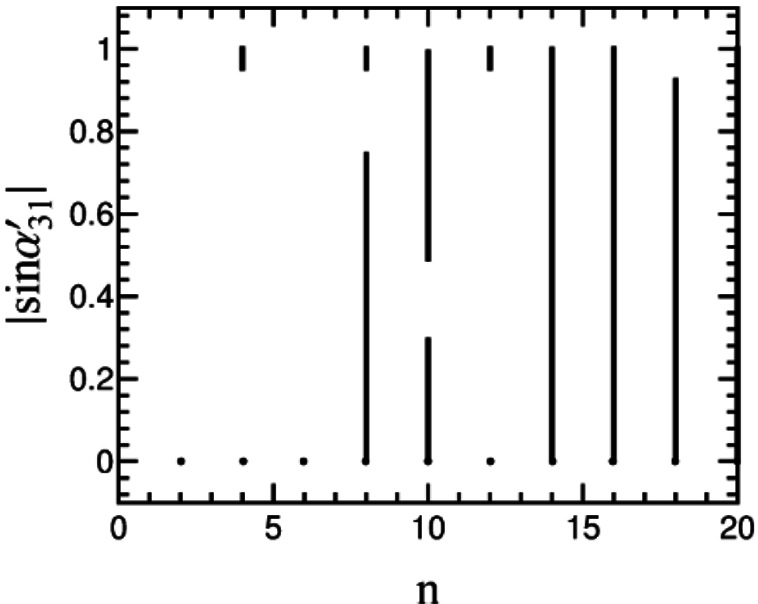}
 \includegraphics[width=0.32\textwidth]{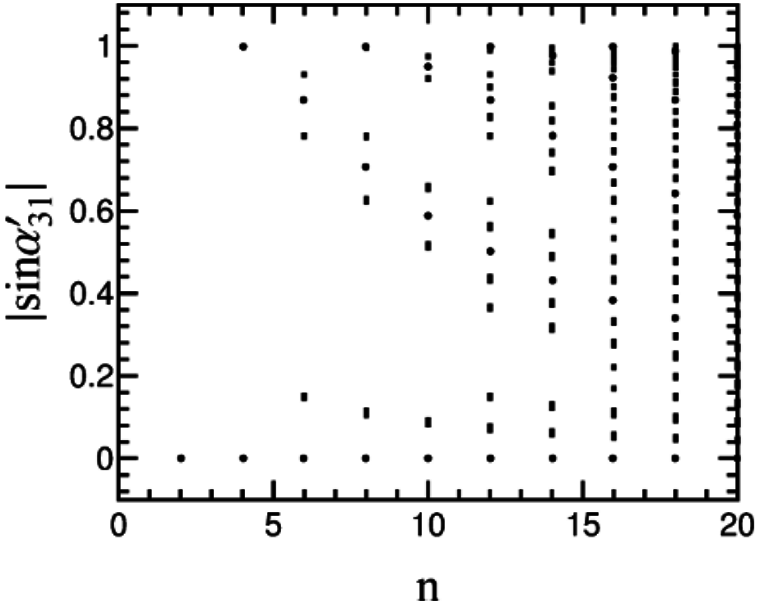}
 \label{semidirect_alpha31}
 \caption{Predictions for $|\sin \alpha'_{31}|$ for the five classes of experimentally viable mixing matrices for semidirect models with a $\Delta(6n^2)$ flavour group. $\alpha'_{31}$ is defined through $\alpha'_{31}=\alpha_{31}-2\delta_\text{CP}$.}
\end{figure}

Out of all possible combinations of subgroups shown in tables 1 and 2 only three different columns are found for $G_\nu\simeq Z_2$ that are not experimentally excluded, while only one distinct row is found for $G_e\simeq Z_2$. Even when combining these columns and rows with a free unitary rotation of the remaining two columns or rows respectively and allowing for arbitrary permutations of the mixing matrices, only five classes of mixing matrices are produced that are not immediately excluded by experiment. The full form of all matrices can be found in \cite{DingORA}. There, the mixing matrices originating from $G_l=\langle ac^sd^t\rangle$ and $G_\nu\simeq Z_2^{bc^xd^x}$ allow for two permutations that produce viable mixing parameters. In figures \ref{semidirect_theta12}-\ref{semidirect_alpha31}, these are shown in the left and middle panel in the top row for each observable where only values of mixing parameters are shown for combinations of group parameters for which every measured observable lies within the experimental limits. The predictions from $G_e=\langle abc^s d^t\rangle$ and $G_\nu=Z_2^{bc^xd^x}$ are shown in the top right panel for each observable, the predictions from $G_e=\langle a c^s d^t\rangle$ and $G_\nu=Z_2^{c^{n/2}}$ in the bottom left panel and the predictions from $G_e=\langle ac^s d^t\rangle$ and $G_\nu=Z_2^{c^{n/2}}$ in the bottom right panel.

In figures \ref{semidirect_theta12}-\ref{semidirect_alpha31}, it can be seen that despite the additional free parameters, the predictions for mixing parameters can often only take discrete or small ranges of values. The predictions are specific, i.e.\ if certain experimental values are measured, first the broad class of mixing matrices can be identified, further the particular $n$ that identifies a group and eventually the unbroken subgroups $G_e$ and $G_\nu$. For very large $n$, the predictions tend to lie increasingly dense until a certain continuous range is filled in the limit $n\rightarrow \infty$. Nevertheless, correlations between parameters that are distinct for the different classes of mixing matrices allow to differentiate them.
\begin{table}
\centering
 \includegraphics[width=0.68\textwidth]{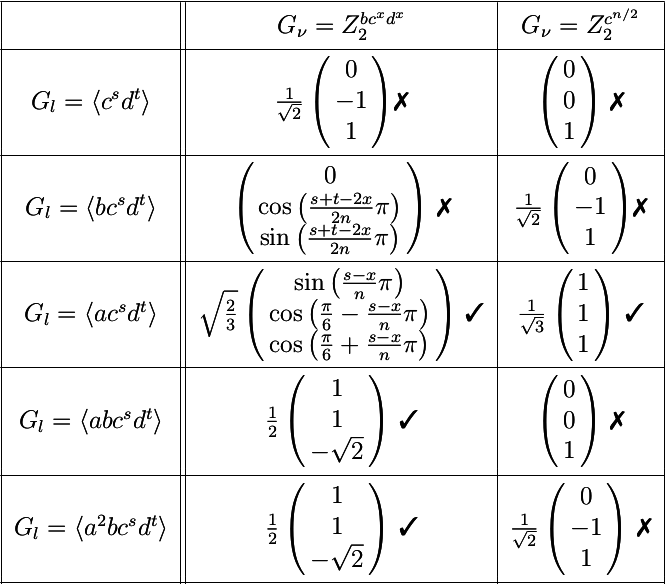}
 \label{neutrinosemidirect}
 \caption{Candidate columns predicted in the neutrino-semidirect case, i.e.\ $G_\nu\simeq Z_2$ for all combinations of subgroups. A tick mark means that this column is viable as one of the columns of the physical mixing matrix, possibly requiring reordering of rows and columns of the mixing matrix. A fail mark indicates that this column is not viable.}
\end{table}

\begin{table}
\centering
 \includegraphics[width=0.68\textwidth]{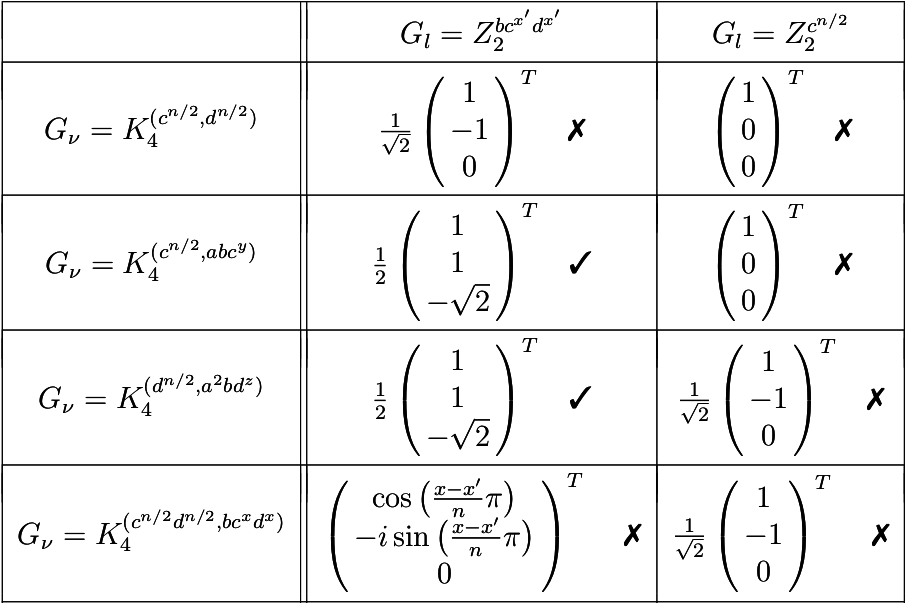}
 \label{leptonsemidirect}
 \caption{Candidate columns predicted in the charged-lepton-semidirect case, i.e.\ $G_l\simeq Z_2$ for all combinations of subgroups. A tick mark means that this column is viable as one of the columns of the physical mixing matrix, possibly requiring reordering of rows and columns of the mixing matrix. A fail mark indicates that this column is not viable.}
\end{table}

\begin{figure}
 \includegraphics[width=0.32\textwidth]{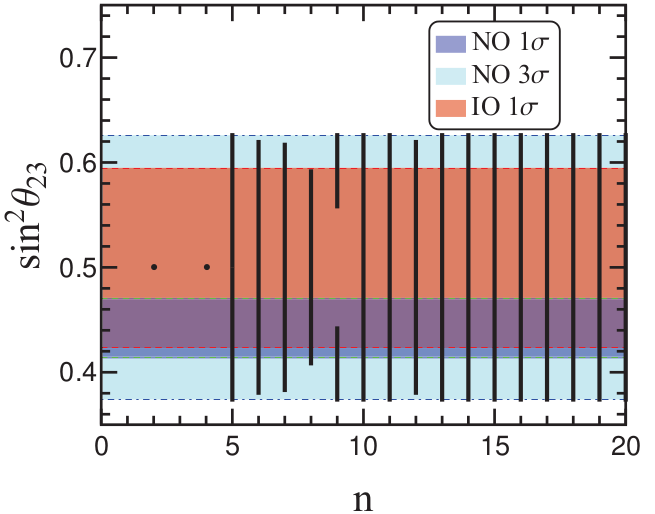}
 \includegraphics[width=0.32\textwidth]{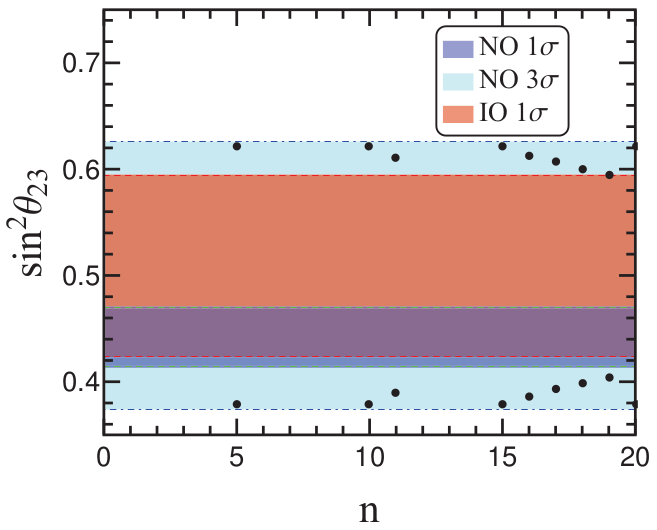}
 \includegraphics[width=0.32\textwidth]{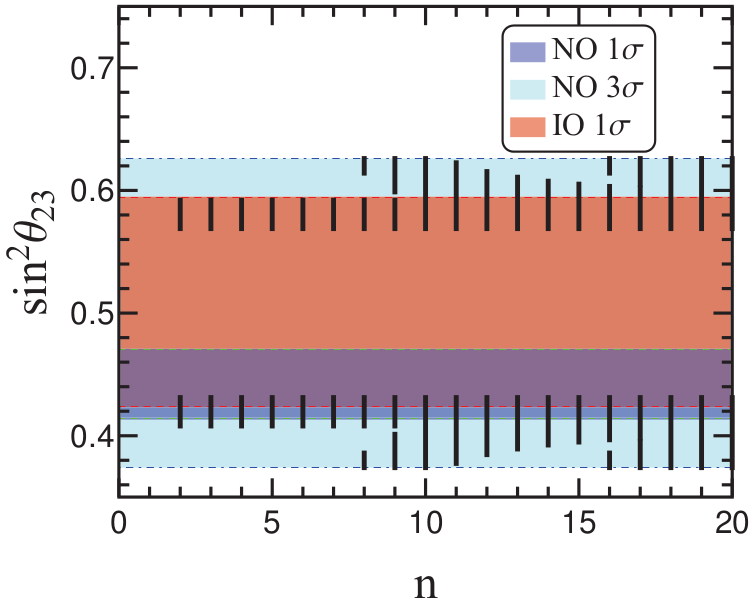}\\
 \includegraphics[width=0.32\textwidth]{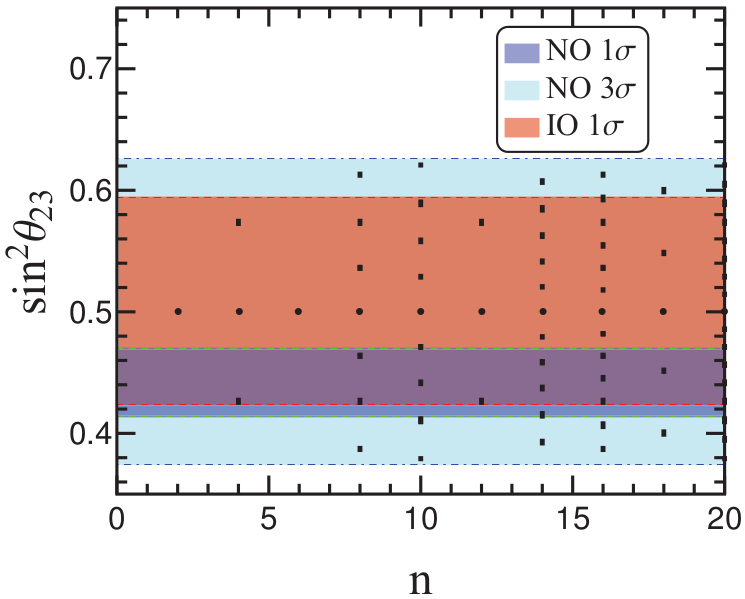}
 \includegraphics[width=0.32\textwidth]{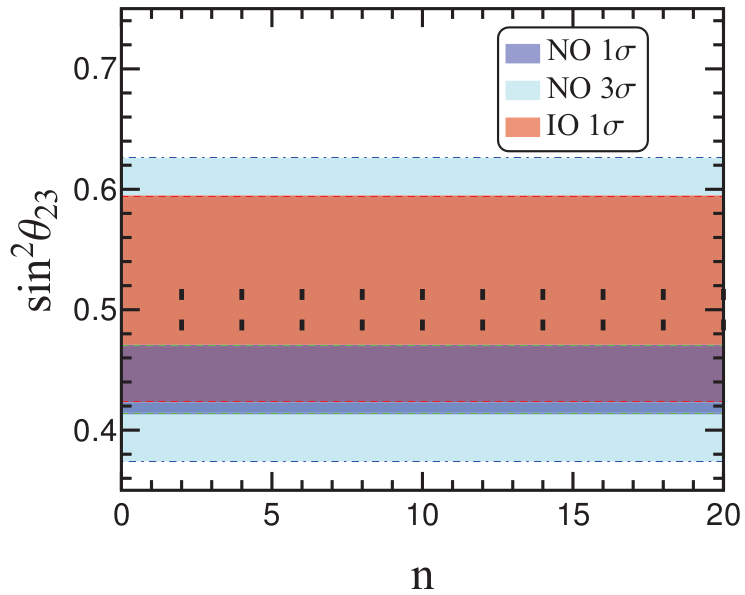}
 \caption{Predictions for $\sin^2\theta_{23}$ for the five classes of experimentally viable mixing matrices for semidirect models with a $\Delta(6n^2)$ flavour group. The coloured bands show the one and three sigma bands respectively.}
 \vspace{1cm}
 \includegraphics[width=0.32\textwidth]{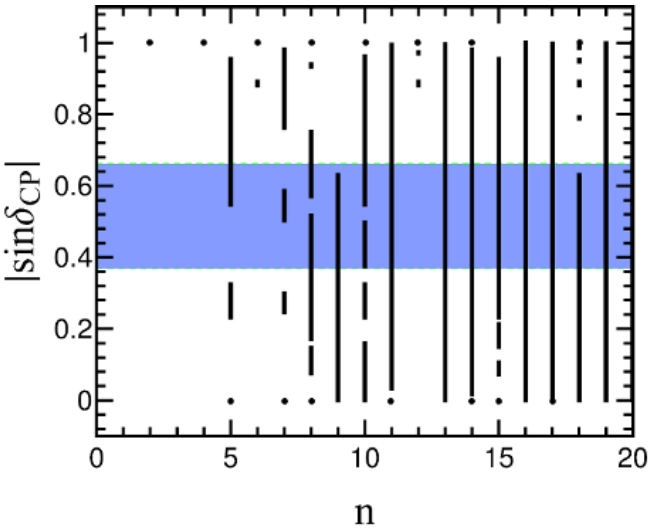}
 \includegraphics[width=0.32\textwidth]{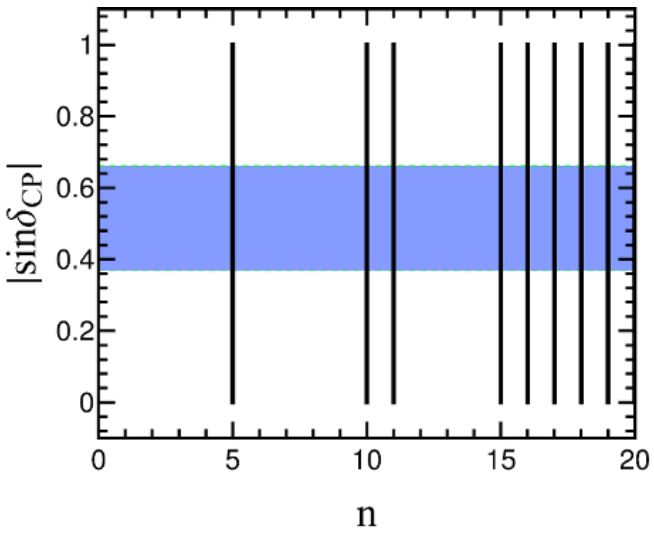}
 \includegraphics[width=0.32\textwidth]{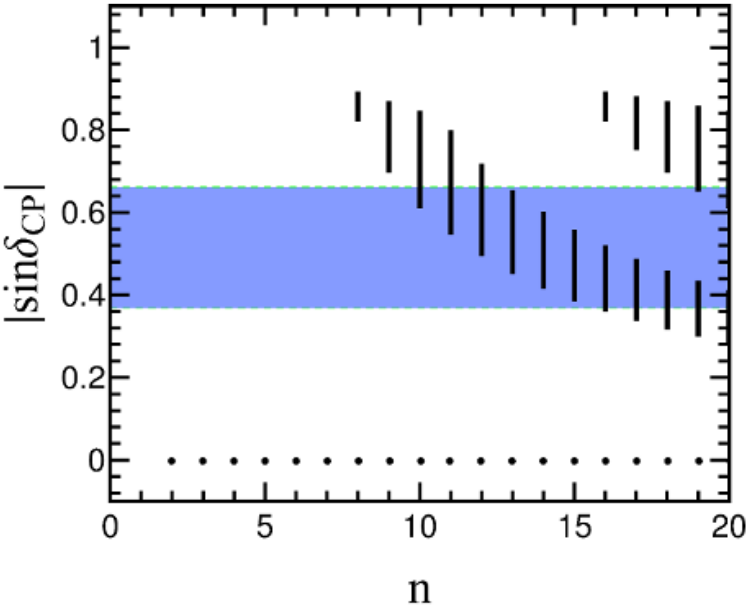}\\
 \includegraphics[width=0.32\textwidth]{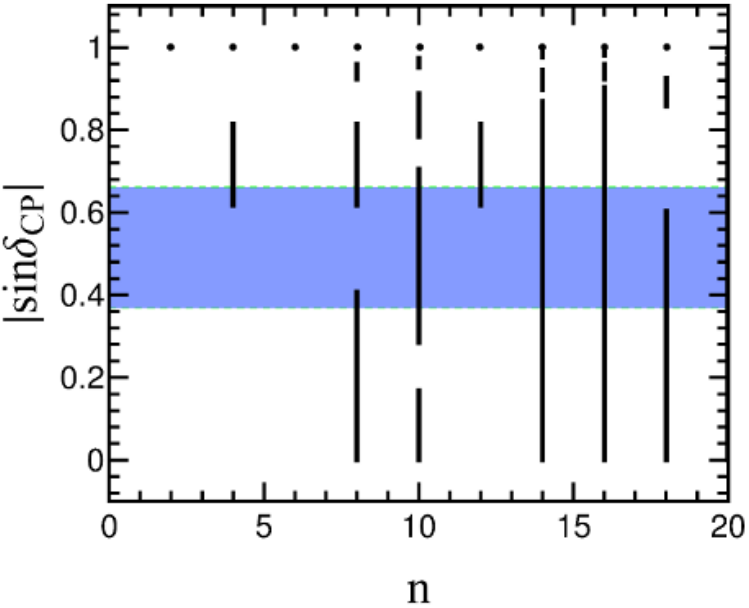}
 \includegraphics[width=0.32\textwidth]{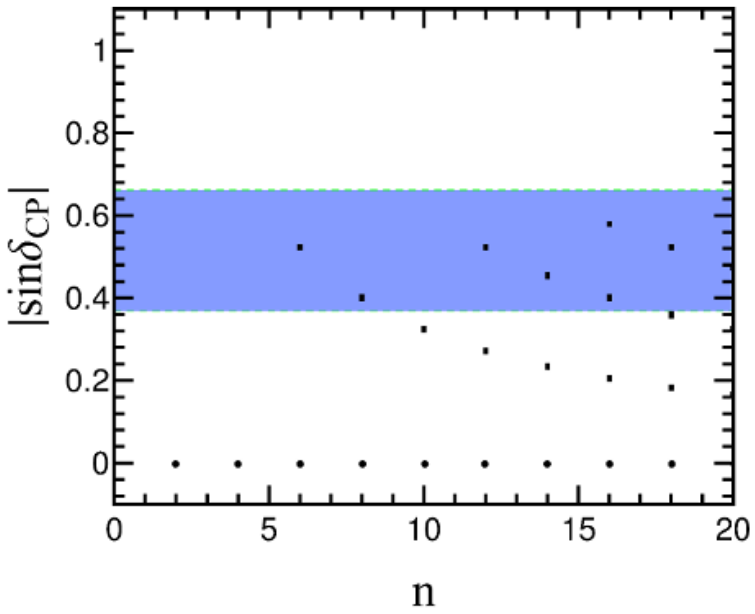}
 \caption{Predictions for $|\sin \delta_\text{CP}|$ for the five classes of experimentally viable mixing matrices for semidirect models with a $\Delta(6n^2)$ flavour group. The band shows the one sigma experimental preference.}
\end{figure}

\section{Conclusions and Future Work}
In this presentation, lepton mixing predictions in direct and semidirect models invariant under $\Delta(6n^2)$ groups and consistent CP transformations have been examined. In direct models, all mixing angles and the Dirac CP phase are predicted purely from symmetry principles. The inclusion of broken invariance under consistent CP transformations additionally allows to predict Majorana phases. Furthermore, only $\Delta(6n^2)$ groups remain viable flavour symmetry groups in direct models. As all members of the $\Delta(6n^2)$ group series can be analysed simultaneously for all $n$, this makes it possible to test the paradigm of direct flavour models with discrete invariance groups generally. In direct models with $\Delta(6n^2)$ groups, discrete values for $\theta_{13}$ are predicted which depend on group parameters. Additionally, predictions that are independent of $n$ can be obtained: The Dirac CP phase is predicted to be zero and the middle column is required to be trimaximal. These predictions are currently compatible with global fits and will be tested experimentally in the near future. 

Furthermore, possible predictions of semidirect models, i.e.\ models in which the flavour group is broken to a $Z_2$ subgroup either for neutrinos or charged leptons are analysed. In this case, despite a variety of possible combinations of subgroups and additional free parameters, discrete or constrained predictions are obtained for all mixing parameters as well as general predictions that allow to differentiate between classes of mixing matrices or to eventually exlude this approach.

In the analyses presented here, no renormalisation corrections to the mixing parameters are taken into account. With increasing experimental accuracy, it might become necessary to consider these.

If experimental results are not excluding models based on $\Delta(6n^2)$ flavour groups, or are even pointing towards specific values of $n$, then the phenomenology of these models is going to be studied in detail, in particular concerning possible breaking mechanisms.

\section{Acknowledgements}
TN would like to thank the organisers of the Discrete2014 conference. TN acknowledges support from the European Union FP7 ITN-INVISIBLES (Marie Curie Actions, PITN- GA-2011- 289442).

\end{document}